\documentclass[a4paper]{jpconf}
\usepackage{graphicx}
\begin{document}

\title{Kinks: Fingerprints of strong electronic correlations}

\author{A. Toschi$^1$, M. Capone$^{2}$, C. Castellani$^{2}$, K. Held$^{1}$}

\address{$^1$ Institut f\"ur Festk\"orperphysik, Technische Universit\"at Wien, Vienna, Austria}
\address{$^2$SMC, CNR-INFM and Dipartimento di Fisica - Universit\`a di Roma ``La Sapienza'', Piazzale Aldo Moro 2, 00185 Roma, Italy}

\ead{held@ifp.tuwien.ac.at}

\begin{abstract}
The textbook knowledge of solid state physics is that the electronic specific heat shows a linear temperature dependence with the leading corrections being a cubic term due to phonons and a cubic-logarithmic term due to
the interaction of electrons with bosons. We have shown that this longstanding conception needs to be supplemented since the generic behavior of the low-temperature electronic specific heat includes a kink if the electrons are sufficiently strongly correlated.
 \end{abstract}

\section{Introduction}
 
Landau's Fermi liquid theory  \cite{Landau} can be considered as the ``standard theory'' of solid state physics. It
predicts that the electronic properties of a (normal) 
metal follow, in the presence of  Coulomb correlations,
 a renormalized (quasiparticle) version of those for 
non-interacting electrons.
 Among the most fundamental physical properties is the 
    heat capacity which, at low temperatures, is 
    predominated by the electronic degrees of freedom. 
     For a normal metal,  Landau's Fermi liquid theory  \cite{Landau}
 predicts a linear
increase of the specific heat capacity with temperature $c_V= \gamma_{FL} T$
and a
cubic term, more precisely a term $\sim T^3 \log(1/T)$, 
as the leading correction \cite{Abrikosov,Ashcroft}.
For free electrons, the prefactor $ \gamma_{0}$ is  directly proportional
to the electronic density of states at the Fermi level 
since, due to the Pauli principle, only these electrons contribute. 
In the case of a Fermi liquid,
we merely need to introduce a quasiparticle renormalization factor
   $0<Z_{\rm FL}<1$ to account for the Coulomb interaction, which enhances
the specific heat of a correlated metal by 
 $\gamma_{FL}=\gamma_0 /Z_{\rm FL}$.
However,  this electronic contribution prevails only at  temperatures much
lower than the Debye temperature. Otherwise the cubic phononic term, which has a much higher prefactor because of its bosonic nature, dominates.

As we have shown in a recent letter \cite{Toschi09}, this
 longstanding conception needs to be supplemented:
For strongly correlated electrons, 
a clear kink in the temperature 
dependence  of the specific heat appears, marking the 
abrupt change from one 
linear behavior to a second one with a reduced slope at 
higher temperatures. This can be shown numerically
solving the Hubbard model \cite{hubbard} within
dynamical mean field theory \cite{dmft}, using 
exact diagonalization as an impurity solver. 
Quantitatively the same results can also been obtained analytically,
following the derivation by Abrikosov {\em et al.}  \cite{Abrikosov}
for the specific heat, taking the recently observed kinks in the self
energy as a starting point \cite{Nekrasov06,Byczuk}.

The experimental confirmation of these theoretical results is
somewhat problematic since typically, at the kink temperature, the phononic contribution to the specific heat is already dominating.
An exception in this respect is
LiV$_2$O$_4$ \cite{LiPaper,Urano}, the first heavy Fermion system with
$d$-electrons, since the kink temperature is here particularly low. And, indeed, the precise inspection of the experimental data
shows a kink in the correct temperature range  \cite{Toschi09}.
In this Proceeding, we will review the previous theoretical results
and include additional data.

\section{Kinks in the self energy}
In the presence of strong electronic correlations, the ${\mathbf k}$-integrated spectral function $A(\omega)$
shows a typical three peak structure with a lower and an upper Hubbard band
and a quasiparticle peak in between. Using the DMFT formula
\begin{equation}
   \Sigma(\omega)=\omega+\mu-1/G(\omega)-\Delta(G(\omega))
    \label{eq:sigmacontributions}
\end{equation}
which relates
Green function and self energy at frequency $\omega$, 
one can show \cite{Byczuk} that such a strongly correlated three peak structure necessarily entails a kink. Note that the hybridization function $\Delta(G(\omega))$
in Eq.\ (\ref{eq:sigmacontributions}) is for a simple semi-circular density of states just 
$\Delta(G(\omega))=\frac{D^2}{4} G(\omega)$ ($D$: half the bandwidth).
For other lattices,  $\Delta(G(\omega))$ depends on the moments
of the non-interacting density of states with a similar term proportional $G(\omega)$ dominating.

The kink argument is now as follows \cite{Byczuk}: The first three terms of the r.h.s of
 Eq.\ (\ref{eq:sigmacontributions})  yield an almost linear frequency dependence for the real part of the self energy over the entire region of the central quasiparticle
peak. The derivative hence gives a quasiparticle renormalization factor
\begin{equation}
Z_{QP}=[1-\partial {\rm Re} \Sigma(\omega)/\partial\omega|_{\omega=0}]^{-1}\label{Eq:Z}
\end{equation}

The last (hybridization) term however yields an additional  contribution to
 ${\rm Re} \Sigma(\omega)$ which basically is proportional to ${\rm Re}G(\omega)$.  This real part can be directly obtained from the spectral function
of the central quasiparticle peak $A(\omega)=-\frac{1}{\pi}{\rm Im} G(\omega)$ via a Kramers-Kronig transformation. {\em Inside} the overall width of the quasiparticle peak, there is a turning point in  $A(\omega)$. Hence,
 ${\rm Re}G(\omega)$ has a linear frequency dependence up to an maximum
at the  turning point $A(\omega)$. After this maximum changes in 
 ${\rm Re}G(\omega)$ are minor. Altogether this means that at low frequencies we have to add the slope obtained from  the $\Delta(G(\omega))$ in
Eq. (\ref{Eq:Z}) yielding an altogether smaller Fermi liquid renormalization
factor $Z_{FL}$ at low energies. In between the two regimes
$Z_{FL}$ and $Z_{QP}$, there is a kink at a frequency $\omega^*$. In summary we hence have:
\begin{equation}
{\rm Re}{\Sigma (\omega)}= \left\{ 
\begin{array}{cc}
(1-\frac{1}{Z_{FL}})\omega & \omega < \omega^* \\
(1-\frac{1}{Z_{QP}})\omega-b & \omega > \omega^*
\end{array}
\label{Eq:Sigma}
\right.
\end{equation}
with the constant $-b=(\frac{1}{Z_{QP}}-\frac{1}{Z_{FL}})\omega^*$
providing for a continuous function.

Before we 
turn to the specific heat in the next section,
let us note that the kink in the self energy directly leads to a kink
in the energy-momentum 
dispersion relation of the correlated electrons.  Let us also remark here, that the result of generic
kinks in the self energy of strongly correlated systems should be robust beyond DMFT, as similar effects are to be expected in cluster \cite{Maier04} 
and diagrammatic extensions \cite{Toschi06a} of DMFT.

\section{Analytical formula for the specific heat}
Based on Eq.\ (\ref{Eq:Sigma}), we have developed an analytical theory for the specific heat
on the basis of a formula by Abrikosov, Gor'kov and Dzyaloshinski (AGD)
for the entropy of a fermionic system at low temperatures\cite{Abrikosov}:
\begin{equation}
S(T) =   \frac{1}{2\pi i}\int_{-\infty}^{\infty} d\epsilon N(\epsilon) \int_{-\infty}^{\infty} \, dy  \, y  \, \frac{ \mbox{e}^y}{ (\mbox{e}^y+1)^2}  [\mbox{log}  G_{R}^{-1}(\epsilon, yT) - \mbox{log} G_{A}^{-1}(\epsilon, yT)],
\label{eq:Abr2}
\end{equation}
 This AGD formula
is  (through a low-temperature diagrammatic expansion)
based on  the low frequency behavior of the self energy $\Sigma(\omega)$ at zero temperature [or the corresponding retarded and advanced
Green function in Eq. (\ref{eq:Abr2})] so that we can directly apply it with  $\Sigma(\omega)$  from Eq.\ (\ref{Eq:Sigma}).
Let us note that $N(\epsilon)$ is the density of states and 
the frequency integral has been rewritten through  a dimensionless variable $y=\omega/T$ ($k_B\equiv 1$). 
The AGD formula  (\ref{eq:Abr2})
can be easily differentiated w.r.t. $T$, so that
the specific heat $c_V(T)  =  T \frac{d S(T)}{d T}$ can be  computed directly. 
With some algebra (see \cite{Toschi09} for more details) and substituting
substituting  $A(\epsilon, yT)$ by  $\delta$-functions, we obtain
the final result
\begin{eqnarray}
c_V(T) &\! =\! &  T \left[\frac{1}{Z_{\rm FL}}\!\!\! \int\limits_{|y|< \frac{\omega^*}{T}}\!\!\!N(\frac{yT}{Z_{\rm FL}}) + \frac{1}{Z_{\rm QP}}\!\!\!\int\limits_{|y|> \frac{\omega^*}{T}}\!\!\! N(\frac{yT}{Z_{\rm QP}}\!+\!b) \right]     d y \,  \frac{y^2  \mbox{e}^y}{ (\mbox{e}^y+1)^2}.
\label{Eq:final}
\end{eqnarray}

Knowing the two renormalization factors $Z_{\rm FL}$ and $Z_{\rm QP}$ and
the kink frequency $\omega^*$ as well as the non-interacting density of states $N(\epsilon)$,  Eq.\ (\ref{Eq:final}) allows us to calculate the specific heat through a simple integral. If vice versa, the specific heat is known, e.g., experimentally, we can employ  Eq.\ (\ref{Eq:final}) as a fit formula for obtaining 
 $Z_{\rm FL}$, $Z_{\rm QP}$ and $\omega^*$.

\section{Experimental validation} 
The latter (fit formula) 
approach had been taken for the experimental validation of the specific heat kinks in LiV$_2$O$_4$ since LDA+DMFT calculations \cite{LDADMFT} for  LiV$_2$O$_4$ \cite{Arita06c} could not provide for a fine enough resolution of the low frequency dependence of the self energy, even when using
the projective quantum Monte Carlo method \cite{Feldbacher04}
which is most appropriate for this purpose.
As Fig.\ 3 of \cite{Toschi06a} shows the three fitting parameters provide for an excellent agreement of Eq.\ (\ref{Eq:final})  and the specific heat of
\cite{LiPaper,Urano}, including a clearly visible {\em kink}. Let us note that we did not subtract a phonon
contribution for the specific heat since (i) it does not alter the existence or
non-existence of a kink, (ii) it is still an order of magnitude  smaller
than the electronic contribution at the kink temperature and (iii)
there is a considerable  arbitrariness depending  on the reference system chosen.

\section{Numerical results}
\begin{figure}[ht!]
\begin{center}
\includegraphics[width=9cm]{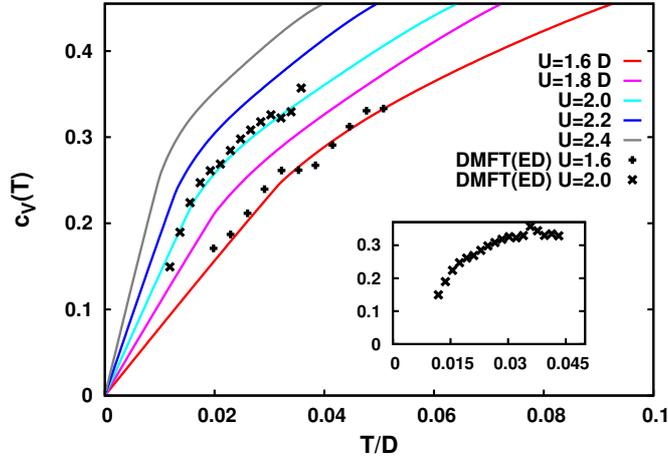}
\end{center}
\caption{Specific heat as a function of temperature for various values of the Coulomb interaction $U$ obtained  via the AGD based equation (\ref{Eq:final}) [solid lines] and numerically by exact diagonalization of the DMFT equations (crosses). Inset: behavior over an extended temperature range which shows a maximum after the kink temperature marking the end of the quasiparticle band. 
\label{Fig:1}}
\end{figure}

Fig.\ \ref{Fig:1} shows results of our final equation  (\ref{Eq:final}) applied to the Hubbard model with a semi-circular density of states with half bandwidth $D\equiv 1$. As an input to Eq.\ (\ref{Eq:final}), $Z_{\rm FL}$, $Z_{\rm QP}$ and $\omega^*$ have been fitted to the
 numerical renormalization data for the DMFT self energy  from Ref.\ \cite{NRGT=0}.
Besides, we have also calculated the DMFT specific heat directly, using
exact diagonalization (ED) with 7 levels in the impurity bath.
Both results agree very well, and  the minor deviations can be attributed to the numerical inaccuracy of the exact diagonalization study for which $c_V$ had to be obtained via a numerical differentiation of the total energy.

In the inset, we show the specific heat over a slightly extended temperature range.
As the quasiparticle peak eventually ends, the specific heat has a maximum after the kink temperature. Note that this maximum is not contained in our formula (\ref{Eq:final}) since it would require an extension of the self energy description
Eq. (\ref{Eq:Sigma}) to higher energies where ${\rm Re} \Sigma(\omega)$ has a
maximum  at the frequencies between Hubbard and quasiparticle bands and  ${\rm Im} \Sigma(\omega)$ becomes important.

\section{Conclusion}
Landau's Fermi liquid theory 
predicts a linear temperature dependence 
of the electronic specific heat capacity with temperature.
We have shown  this longstanding conception needs to be supplemented:
For strongly correlated electrons, 
a clear kink in the temperature 
dependence  appears, marking the abrupt change from one 
linear behavior to a second one with a reduced slope at 
higher temperatures. Recent experiments on 
LiV$_2$O$_4$,
an ideal material for studying the electronic specific heat,
confirm our theory. A consequence of our findings is that materials
with correlated electrons 
are more resistive to cooling at low temperatures (where cooling is particularly difficult)    
than expected  from the behavior at intermediate temperatures.

\ack 
We thank R. Bulla and M. Nohara for making available their raw data.
This work has been supported in part
by the European-Indian cooperative network MONAMI.

\end{document}